\documentclass[acmsmall]{acmart}   
\usepackage[capitalise,nameinlink]{cleveref}
\usepackage{enumitem}
\usepackage{multirow}

\usepackage{listings}

\usepackage{subcaption}

\crefformat{section}{#2\S{}#1#3}
\crefname{lstlisting}{Listing}{Listings}

\definecolor{SoftBlue}{HTML}{005CC5}
\definecolor{WarmGray}{HTML}{6A737D}
\definecolor{DarkRed}{HTML}{D73A49}
\definecolor{WarmGray}{HTML}{6A737D}
\definecolor{Teal}{HTML}{22863A}
\definecolor{BurntOrange}{HTML}{E36209}
\definecolor{backcolor}{rgb}{0.95,0.95,0.95}
\definecolor{shadecolor}{rgb}{0.95,0.95,0.95}

\usepackage{framed}
\newenvironment{lstshade}%
  {\setlength{\fboxsep}{1pt}\vspace{1pt}\begin{snugshade}}%
  {\end{snugshade}\vspace{2pt}}
\BeforeBeginEnvironment{lstlisting}{\begin{lstshade}}
\AfterEndEnvironment{lstlisting}{\end{lstshade}}

\colorlet{DSLKeyword}{SoftBlue}
\colorlet{DSLStatement}{Teal}
\colorlet{DSLComment}{WarmGray}
\colorlet{DSLSymbol}{DarkRed}

\lstdefinelanguage{Rust}{
  keywords={as,break,const,continue,crate,else,enum,extern,false,fn,for,if,
  impl,in,let,loop,match,mod,move,mut,pub,ref,return,self,Self,static,struct,
  super,trait,true,type,unsafe,use,where,while,async,await,dyn,abstract,become,
  box,do,final,macro,override,priv,typeof,unsized,virtual,yield,try,union,raw},
  morecomment=[l]{//},
  morecomment=[s]{/*}{*/},
}

\lstdefinestyle{mystyle}{
  aboveskip=0pt,
  belowskip=0pt,
  abovecaptionskip=3pt,
  belowcaptionskip=0pt,
  keywordstyle=\color{DSLKeyword},
  commentstyle=\color{DSLComment},
  basicstyle=\linespread{0.8}\ttfamily\footnotesize,
  breakatwhitespace=false,
  breaklines=true,
  captionpos=b,
  keepspaces=true,
  numbers=left,
  numbersep=1pt,
  numberstyle=\tiny,
  showspaces=false,
  showstringspaces=false,
  escapechar=$,
  language=Rust,
}
\newcommand{\C}[1]{{\small\tt#1}}
\newcommand{\rustc}[0]{\C{rustc}}
\newcommand{\topspace}{\rule{0pt}{2.0ex}}

\setcopyright{cc}
\setcctype{by}
\acmDOI{10.1145/3832220}
\acmYear{2026}
\acmJournal{PACMSE}
\acmVolume{3}
\acmNumber{ISSTA}
\acmArticle{ISSTA129}
\acmMonth{10}
\acmSubmissionID{issta26main-p1206-p}
\received{2026-01-30}
\received[accepted]{2026-06-25}

\begin{document}

\title{Rust's Type Checker Implementation Is Unsound: An Empirical Study on Soundness Bugs in rustc}

\author{Yusung Sim}
\orcid{0000-0003-3641-593X}
\affiliation{%
  \institution{KAIST}
  \city{Daejeon}
  \country{Republic of Korea}
}
\email{yusungsim@kaist.ac.kr}

\author{Sukyoung Ryu}
\orcid{0000-0002-0019-9772}
\affiliation{%
  \institution{KAIST}
  \city{Daejeon}
  \country{Republic of Korea}
}
\email{sryu.cs@kaist.ac.kr}

\author{Jaemin Hong}
\orcid{0000-0003-4067-7369}
\affiliation{%
  \institution{UNIST}
  \city{Ulsan}
  \country{Republic of Korea}
}
\email{jaemin.hong@unist.ac.kr}
\affiliation{%
  \institution{Yale University}
  \city{New Haven}
  \country{USA}
}
\email{jaemin.hong@yale.edu}

\begin{abstract}
  Rust is claimed to be a \emph{type-sound} language capable of preventing
  various undesirable behaviors, including memory bugs.
  However, \rustc{}, the official Rust compiler, is not immune to defects;
  it contains \emph{soundness bugs}, where the compiler accepts programs that
  should be rejected during type checking.
  In this work, we present an empirical study of 30 issues that report potential
  soundness bugs in \rustc{}, collected from the GitHub issue tracker between
  January 1, 2022 and September 1, 2025.
  We analyze each issue in depth, focusing on its \emph{affected feature},
  \emph{symptom} (how the feature is mishandled), \emph{consequence} (the
  resulting undesirable behavior), \emph{triggering features}, \emph{community
  consensus} regarding whether it is a bug, and \emph{lifecycle}, including
  introduction, discovery, and fix.
  Furthermore, we investigate existing artifacts, including implementations such
  as AddressSanitizer, Miri, Chalk, and a-mir-formality, alongside documentation
  such as the Rust Reference, the FLS, and Rust RFCs to assess their potential
  as oracles for testing the type soundness of \rustc{}.
  Our key findings indicate that:
  (1) Certain soundness bugs, typically triggered by \emph{implied bounds} or
  \emph{trait objects}, compromise memory safety.
  (2) Sound type checking is challenged by edge cases involving \emph{associated
  types} and the interaction between \emph{lifetimes} and \emph{traits}.
  (3) Most bugs persist from the initial introduction of the relevant features
  and require significant time to be discovered.
  (4) While AddressSanitizer and Miri can detect soundness bugs that lead to
  memory bugs, a-mir-formality and Chalk are currently immature despite their
  potential to identify other bug categories.
  (5) Existing documentation frequently fails to provide precise explanations of
  the language semantics.
\end{abstract}

\ccsdesc[500]{Software and its engineering~Compilers}
\ccsdesc[500]{Software and its engineering~Software defect analysis}
\ccsdesc[300]{Software and its engineering~Empirical software validation}
\ccsdesc[300]{Theory of computation~Type theory}

\keywords{Rust, type system, compiler bugs, empirical study, rustc}

\maketitle

\section{Introduction}
\label{sec:intro}
Rust~\cite{matsakis2014rust} is designed to be a \emph{type-sound} language,
aiming to prevent various undesirable behaviors, including memory bugs like
use-after-free, at compile time through type checking~\cite{jung2017rustbelt}.
Recognizing this benefit, researchers and practitioners have adopted Rust for
systems programming to enhance software reliability.
Operating systems, web browsers, and network stacks have been developed in
Rust~\cite{levy2017multiprogramming, lankes2020rustyhermit,
narayanan2020redleaf, boos2020theseus, hu2024unishyper, peng2024framekernel,
dai2024verifying, hong2024taming, chiang2024securing, anderson2016engineering,
chen2025atmosphere, shang2025lwrustip};
legacy systems, including Linux, have begun integrating Rust~\cite{li2024rust,
netstack3, rust-dropbox, rust-coreutils};
and techniques to translate legacy code into Rust have been extensively
studied~\cite{emre2021translating, emre2023aliasing, zhang2023ownership,
hong2023concrat, hong2024dont, hong2024tag, wu2025genc2rust, hong2025forcrat,
hong2025automatically, hong2024type, zhou2025c2rusttv, xu2025optimizing,
zhang2025systematic, cai2026rustmap, zhang2025scalable}.
All these lines of work rely on the fundamental assumption that the Rust type
checker is sound.

Unfortunately, \rustc{}, the official Rust compiler, is not free from defects.
Liu et al.~\cite{liu2025empiricala} recently demonstrated that the compiler
suffers from various bugs, including crashes, miscompilations, and improper
diagnostic messages.
While all bugs warrant attention, \emph{soundness bugs}, where the type checker
accepts a program that should be rejected, are especially critical, as they
undermine Rust's core value of type soundness guarantee.
Notably, among the 301 \rustc{} bugs collected by Liu et al., 22 were identified
as soundness bugs.

Despite their importance, soundness bugs in the \rustc{} implementation have not
yet been studied in depth.
For instance, it remains unclear which language features are most prone to
soundness bugs, even though such insights would allow researchers to prioritize
specific features when developing testing techniques or formalizing the type
system.
While Liu et al. analyzed \rustc{} bugs and provided valuable insights, their
study encompassed all bug categories rather than focusing specifically on
soundness bugs.
We believe that an in-depth study dedicated to soundness bugs can benefit the
community and complement existing work.

In this work, we study 30 \emph{soundness issues} collected from the \rustc{}
GitHub issue tracker, each representing a case where a developer reported that
the compiler accepted code that should not pass type checking.
We analyze each issue in depth to characterize the following:
the \emph{affected feature},
the \emph{symptom} (i.e., how the feature is erroneously handled),
the \emph{consequence} (i.e., the specific undesirable behavior enabled by the
issue),
the features that \emph{trigger} the issue,
the \emph{consensus} of the community regarding whether it is indeed a bug,
and the \emph{lifecycle} of the issue, encompassing its introduction, discovery,
and fix.

Furthermore, we investigate artifacts that can aid in testing the type soundness
of \rustc{}.
Liu et al.'s study showed that existing testing tools for
\rustc{}~\cite{fuzz-rustc, tree-splicer, icemaker, sharma2023rustsmith,
yang2024rusttwins, wang2024rustlantis, dewey2015clp} revealed crash bugs but
failed to identify a single soundness bug.
A primary reason for this is the absence of a \emph{test oracle} capable of
determining whether a given program should be accepted or rejected.
To address this, RustSmith~\cite{sharma2023rustsmith} employs differential
testing, comparing results across different \rustc{} versions and
configurations.
However, its effectiveness is limited because most soundness issues persist
across multiple versions and settings, as we observe in our study.
This underscores the necessity of a test oracle tailored to type soundness.

For this reason, we study the following implementations, which serve as
potential test oracles:
AddressSanitizer~\cite{serebryany2012addresssanitizer} (a memory bug detector),
Miri~\cite{jung2026miri, gh-miri} (a Rust interpreter for detecting undefined
behavior),
Chalk~\cite{gh-chalk} (an implementation of the Rust trait system based on a
logic solver),
and a-mir-formality~\cite{gh-amirformality} (an executable model of the Rust
type system).
We assess their effectiveness by checking whether they can determine that the
programs from the collected issues should be rejected.

In addition, we study existing documentation that explains the semantics of
Rust.
Previous research has demonstrated that \emph{mechanized
specifications}~\cite{park2021jiset, youn2024bringing} can serve as effective
test oracles for language implementations,
detecting numerous bugs in JavaScript engines and
transpilers~\cite{park2021jest, park2023feature, ryu2024javascript} and
WebAssembly engines~\cite{youn2025west}, as well as soundness bugs in the P4
type checker~\cite{lee2026failing}.
Although Rust currently lacks a full specification~\cite{are-we-spec-yet},
documents such as the Rust Reference~\cite{rust-reference}, the FLS~\cite{fls},
and Rust RFCs~\cite{rust-rfcs} provide explanations of the semantics.
Therefore, we investigate the current state of this documentation and assess
whether these explanations are sufficiently clear to develop test oracles for
type soundness.


\vspace{0.5em} {\bf Contributions.} Overall, our contributions are as follows:
\begin{itemize}[leftmargin=*]
  \item We construct a three-and-a-half-year dataset of 30 soundness issues in
    \rustc{} (\cref{sec:method}).
  \item We conduct an empirical study of the soundness issues across multiple
    dimensions, including their affected features, symptoms, consequences,
    triggers, consensus, and lifecycles (\cref{sec:bugs}).
  \item Based on our analysis, we summarize key findings and provide actionable
    suggestions for researchers and \rustc{} developers to better detect and
    mitigate soundness bugs (\cref{sec:findings}).
  \item We assess existing artifacts that could potentially serve as oracles for
    testing type soundness, discussing their current capabilities and providing
    guidelines for their use (\cref{sec:oracles}).
\end{itemize}

{\bf Summary of Findings.} We summarize our key findings as follows:
\begin{itemize}[leftmargin=*]
  \item Certain soundness bugs lead to memory bugs; specifically, \emph{implied
    bounds} and \emph{trait objects} are critical triggering features that can
    compromise memory safety.
  \item \emph{Associated types} frequently serve as a mechanism for exploiting
    edge cases in other features, and the interaction between \emph{lifetimes}
    and \emph{traits} poses challenges for sound type checking.
  \item Many bugs have persisted since the initial introduction of the relevant
    features, typically requiring a long time to discover and remaining present
    across multiple compiler versions.
  \item AddressSanitizer and Miri can detect soundness bugs that manifest as
    memory bugs; other bug categories could be identified by a-mir-formality and
    Chalk in the future, but they are currently immature.
  \item Existing documents are often unclear, and constructing a mechanized
    specification for Rust cannot be achieved simply by collecting them.
\end{itemize}

\section{Background: Language Features of Rust}
\label{sec:background}
In this section, we provide a brief overview of the language features of Rust
that commonly appear in the soundness issues we collected.

\subsection{Lifetimes}
\label{sec:background:lifetime}

In Rust, a \emph{reference} is a pointer with a static guarantee that it points
to a valid value.
Each reference has a \emph{lifetime}, which specifies the duration for which the
reference is valid.
Lifetimes are key to ensure memory safety;
for instance, the compiler ensures that a lifetime ends before the referenced
value goes out of scope.
A reference has a type \C{\&'a T}, where \C{'a} is the lifetime and \C{T} is the
type of the value.

Functions can have \emph{lifetime parameters} to express the relationship
between the lifetimes of the parameters and the return value.
For example, the following function with two lifetime parameters \C{'a} and
\C{'b} takes two references to integers (\C{i32}) and returns the first one,
specifying that the lifetime of the return value is the same as that of the
first parameter:

\vspace{-0.5em}
\begin{lstlisting}[basicstyle=\linespread{0.85}\ttfamily\footnotesize,]
fn foo<'a, 'b>(v: &'a i32, u: &'b i32) -> &'a i32 { v }
\end{lstlisting}
\vspace{-0.5em}

When a lifetime \C{'a} is longer than or equal to another lifetime \C{'b}, i.e.,
\C{'a} \emph{outlives} \C{'b}, a reference of lifetime \C{'a} can be used where
a reference of lifetime \C{'b} is expected.
By default, different lifetime parameters are not considered to outlive each
other, but programmers can introduce \emph{lifetime bounds} of the form \C{'a:
'b}, implying \C{'a} outlives \C{'b}, as shown below:

\vspace{-0.5em}
\begin{lstlisting}[basicstyle=\linespread{0.85}\ttfamily\footnotesize,]
fn foo<'a, 'b>(v: &'a i32) -> &'b i32 { v } // error
fn bar<'a: 'b, 'b>(v: &'a i32) -> &'b i32 { v } // ok
fn baz() { let p; { let v = 0; p = bar(&v); } *p; } // error
fn qux() { let v = 0; let p = bar(&v); *p; } // ok
\end{lstlisting}
\vspace{-0.5em}

\noindent
In \C{foo}, returning \C{v} is disallowed because there is no guarantee that
\C{'a} outlives \C{'b};
in \C{bar}, \C{'a: 'b} allows it.
The compiler checks whether the specified lifetime bounds are satisfied at each
call site:
\C{baz} is rejected because \C{p} is used after \C{v} goes out of scope, while
\C{qux} is accepted.




Since a value must remain valid while its reference is live, a type \C{\&'a T}
is considered \emph{well-formed} only when every lifetime appearing in \C{T}
outlives \C{'a}.
For example, \C{\&'a \&'b i32} is well-formed only if \C{'b: 'a} holds.

To reduce the burden of stating lifetime bounds, Rust supports \emph{implied
bounds}, which automatically introduce lifetime bounds deduced from the
well-formedness of the types used in each definition (e.g., the parameter and
return types of a function).
In the following example, \C{foo} passes type checking due to the implied bound
\C{'b: 'a} and is equivalent to \C{bar}:

\vspace{-0.5em}
\begin{lstlisting}[basicstyle=\linespread{0.85}\ttfamily\footnotesize,]
fn foo<'a, 'b>(v: &'a &'b i32) {}
fn bar<'a, 'b: 'a>(v: &'a &'b i32) {}
\end{lstlisting}
\vspace{-0.5em}

\noindent
Note that implied bounds are distinct from Rust's lifetime elision, which allows
programmers to omit explicit lifetime annotations (e.g., writing \C{\&i32}
instead of \C{\&'a i32}), not lifetime bounds, which the compiler then infers
based on fixed syntactic patterns.

So far, the names of lifetimes have been explicitly introduced by lifetime
parameters, but there is one exception:
the special lifetime \C{'static} denotes the entire duration of the program and
does not need to be defined as a lifetime parameter.
By definition, \C{'static} outlives any lifetime.
A reference to a global variable or a constant has the \C{'static} lifetime.


\subsection{Traits}
\label{sec:background:trait}

A \emph{trait} represents a set of types, similar to abstract classes or
interfaces in object-oriented languages and type classes in Haskell.
Consider the following example:

\vspace{-0.5em}
\begin{lstlisting}[basicstyle=\linespread{0.85}\ttfamily\footnotesize,]
trait X { fn foo(self); } $\label{line:trait:1}$
struct S; impl X for S { fn foo(self) { println!("S"); } } $\label{line:trait:2}$
fn bar<T: X>(v: T) { v.foo(); } $\label{line:trait:3}$
fn baz<T>(v: T) where T: X { v.foo(); } $\label{line:trait:4}$
fn qux(v: impl X) { v.foo(); } $\label{line:trait:5}$
\end{lstlisting}
\vspace{-0.5em}

\noindent
Here, \C{X} denotes a set of types that implement the method \C{foo}
(line~\ref{line:trait:1}), and \C{S} belongs to \C{X}, i.e., \C{S}
\emph{implements} \C{X} (line~\ref{line:trait:2}).
The function \C{bar} takes a value of any type \C{T} that implements \C{X} and
calls \C{foo} (line~\ref{line:trait:3}).
In \C{bar}, \C{X} is a \emph{trait bound} of \C{T}.
Trait bounds can also be specified in a \C{where} clause
(line~\ref{line:trait:4}).
In addition, instead of introducing a type parameter and its bound, we can use
an \emph{impl trait} type, such as \C{impl X}, as a parameter type
(line~\ref{line:trait:5}).
This use of an \C{impl} trait type is called an \emph{argument-position impl
trait} (APIT).
All the functions \C{bar}, \C{baz}, and \C{qux} are equivalent, representing the
same set of possible arguments with different syntax.
%

It is possible to implement a trait for multiple types using a single \C{impl},
as shown below:

\vspace{-0.5em}
\begin{lstlisting}
struct S; struct R; struct Q; trait X {} impl X for S {} impl X for R {} $\label{line:blanket1}$
trait Y {} impl<T: X> Y for T {} trait Z {} impl<T> Z for T {} $\label{line:blanket2}$
\end{lstlisting}
\vspace{-0.5em}

\noindent
\C{S} and \C{R} implement \C{X}, but \C{Q} does not (line~\ref{line:blanket1}).
Since \C{Y} is implemented for all types that implement \C{X}, both \C{S} and
\C{R} implement \C{Y}, but \C{Q} does not;
on the other hand, \C{Z} is implemented for all types, including \C{S}, \C{R},
and \C{Q} (line~\ref{line:blanket2}).

Traits can require types not only to provide particular methods but also to
specify \emph{associated types}, as in the following example:

\vspace{-0.5em}
\begin{lstlisting}
trait X { type A; fn foo(self) -> Self::A; } $\label{line:assoc:1}$
struct S; impl X for S { type A = i32; fn foo(self) -> Self::A { 0 } } $\label{line:assoc:2}$
struct R; impl X for R { type A = f32; fn foo(self) -> Self::A { 0.0 } } $\label{line:assoc:3}$
fn bar<T: X>(v: T) -> T::A { v.foo() } $\label{line:assoc:4}$
fn baz<T: X>(v: T) -> <T as X>::A { v.foo() } $\label{line:assoc:5}$
fn qux<T: X<A = i32>>(v: T) -> i32 { v.foo() } $\label{line:assoc:6}$
\end{lstlisting}
\vspace{-0.5em}

\noindent
A type implementing \C{X} must specify an associated type \C{A}, and its method
\C{foo} must return a value of type \C{A} (line~\ref{line:assoc:1}).
Both \C{S} and \C{R} implement \C{X}, defining different types for \C{A}
(lines~\ref{line:assoc:2} and \ref{line:assoc:3}).
The function \C{bar} returns the value from \C{foo}, whose type is denoted
\C{T::A} (line~\ref{line:assoc:4}).
We can also write \C{<T as X>::A} to disambiguate which trait declares the
associated type (line~\ref{line:assoc:5}), which is useful when multiple traits
define an associated type with the same name.
Additionally, to specify a set of types implementing a trait with a specific
associated type, we can use a trait bound like \C{X<A = i32>};
this allows \C{S} to be passed to \C{qux}, but not \C{R}
(line~\ref{line:assoc:6}).

Traits are also used to hide the concrete return type of a function by
specifying the return type as \C{impl X}, called a \emph{return-position impl
trait} (RPIT).
Consider the example below:

\vspace{-0.5em}
\begin{lstlisting}
trait X { fn foo(self); } $\label{line:rpit:1}$
struct S; impl X for S { fn foo(self) {} } impl S { fn bar(self) {} } $\label{line:rpit:2}$
fn baz() -> impl X { S } $\label{line:rpit:3}$
fn qux() { let v = baz(); v.foo(); /* ok */ v.bar(); /* error */ } $\label{line:rpit:4}$
\end{lstlisting}
\vspace{-0.5em}

\noindent
\C{S} defines two methods \C{foo} and \C{bar}, while only \C{foo} is part of the
trait \C{X} (lines~\ref{line:rpit:1} and \ref{line:rpit:2}).
The function \C{baz} returns \C{S}, but it hides this information from callers,
revealing only that the return type implements \C{X} (line~\ref{line:rpit:3}).
Therefore, \C{qux} can call \C{foo} on the returned value, but not \C{bar}
(line~\ref{line:rpit:4}).
Note that the purpose of an RPIT is not to allow returning values of different
types.
Even when using an RPIT, a single underlying return type must exist.
For instance, the following code is rejected even though both \C{S} and \C{R}
implement \C{X}:

\vspace{-0.5em}
\begin{lstlisting}
trait X {} struct S; impl X for S {} struct R; impl X for R {}
fn foo(v: i32) -> impl X { if v == 0 { S } else { R } } // error
\end{lstlisting}
\vspace{-0.5em}

To enable returning values of different types, Rust supports \emph{trait
objects}, which accompany values with virtual tables for dynamic dispatch.
The type of a trait object is \C{dyn X}, where \C{X} is the implemented trait,
as in the following example:

\vspace{-0.5em}
\begin{lstlisting}
fn foo(v: i32) -> &'static dyn X { if v == 0 { &S } else { &R } } // ok
\end{lstlisting}
\vspace{-0.5em}

\noindent
Note that a trait object is \emph{dynamically sized}, i.e., its size is unknown
at compile time, because values of different types can be stored.
It is thus typically placed behind a pointer.

\subsection{Higher-Rank Polymorphism}
\label{sec:background:hr}

Rust supports rank-1 type polymorphism (e.g., generic functions), but not
higher-rank type polymorphism (e.g., functions that take generic functions).
However, it supports higher-rank \emph{lifetime} polymorphism because it is
extremely useful when dealing with function pointers:

\vspace{-0.5em}
\begin{lstlisting}
fn foo<'a>(v: &'a i32) -> i32 { *v }$\label{line:hr:1}$
fn bar<'a>(f: fn(&'a i32)->i32) {{let v=0;f(&v);} {let u=1;f(&u);}} // error$\label{line:hr:2}$
fn baz(f: for<'a> fn(&'a i32)->i32) {{let v=0;f(&v);} {let u=1;f(&u);}}// ok$\label{line:hr:3}$
\end{lstlisting}
\vspace{-0.5em}

\noindent
The function \C{foo} takes a reference of any lifetime (line~\ref{line:hr:1}).
Another function may want to take a function pointer to \C{foo} and call it on
references with different lifetimes.
This cannot be achieved with rank-1 polymorphism:
\C{bar} takes a function that accepts a reference of \emph{some} lifetime,
failing to pass both \C{\&v} and \C{\&u} (line~\ref{line:hr:2}).
With higher-rank polymorphism, \C{baz} takes a function that accepts a reference
of \emph{any} lifetime, allowing both calls (line~\ref{line:hr:3}).

Higher-rank polymorphism is also useful when reference types implement traits:

\vspace{-0.5em}
\begin{lstlisting}
trait X { fn foo(self); }$\label{line:hrtb:1}$
fn bar<T>(mut v: T) where for<'a> &'a T: X {$\label{line:hrtb:2}$
  { let u = v; (&u).foo(); v = u; } { let w = v; (&w).foo(); } }$\label{line:hrtb:3}$
struct S; impl<'a> X for &'a S { fn foo(self) {} } fn baz() { bar(S); }$\label{line:hrtb:4}$
\end{lstlisting}
\vspace{-0.5em}

\noindent
Using a \emph{higher-rank trait bound} (HRTB), the function \C{bar} requires the
type parameter \C{T} to satisfy the condition that \C{\&'a T} implements \C{X}
for any lifetime \C{'a} (line~\ref{line:hrtb:2}).
This allows the function body to call the method \C{foo}, which is provided by \C{X},
using references of different lifetimes (line~\ref{line:hrtb:3}).
Since \C{\&'a S} implements \C{X} for any lifetime \C{'a}, \C{baz} can pass
\C{S} to \C{bar} (line~\ref{line:hrtb:4}).

\section{Collecting Soundness Issues}
\label{sec:method}
We collected \rustc{}'s soundness issues reported between January 1, 2022, and
September 1, 2025, focusing on issues pertinent to recent compiler releases.
This timeframe aligns with the two most recent \textit{editions} (boundaries for
introducing backward-incompatible changes) of the language:
Rust 2021, released on October 21, 2021, and Rust 2024, released on February 20,
2025.
Consequently, our target timeframe encompasses over three and a half years,
corresponding to the active use of the Rust 2021 and 2024 editions.

%

Since the issue tracker of the official GitHub repository for \rustc{} contains
various kinds of issues, we filtered for soundness issues through automatic
filtering based on the labels assigned to each issue and manual inspection.
We initially collected issues related to type checking by utilizing the area
labels (prefixed with ``A-''), which indicate the issue-relevant compiler area,
component, or language feature.
Specifically, we retained issues with the fourteen labels classified as
type-related ones by previous work~\cite{liu2025empiricala}:
A-type-system, A-inference, A-closures, A-coercions, A-const-generics, A-DSTs,
A-zst, A-trait-system, A-impl-trait, A-trait-objects, A-auto-traits,
A-implied-bounds, A-coinduction, and A-coherence.
This yielded 969 issues.

%
%

Since this set still includes issues other than soundness (e.g., bugs in
documentation or non-bugs like feature proposals), we further refined the
dataset using two criteria.
First, we kept only issues marked with either C-bug or I-unsound, which
signifies that the issue is a bug report or related to type soundness.
Second, we filtered out issues marked with specific labels indicating
irrelevance to soundness, which we categorize as follows:
\begin{itemize}[leftmargin=*]
  \item Documentation and tool issues: A-docs, A-rustdoc, A-rustdoc-json,
    A-rustdoc-ui, T-rustdoc, and A-clippy, which represent wrong documentation
    or issues in external tools.
  \item Diagnostics issues: A-diagnostics, A-suggestion-diagnostics,
    NLL-diagnostics, and const-generics-bad-diagnostics, which are pertinent to
    the quality of error messages.
  \item Alternative symptoms: I-crash, I-ICE, I-hang, I-compilemem,
    I-compiletime, and A-codegen, which represent symptoms like compiler crashes
    or performance issues.
  \item Non-bugs: C-tracking-issue, C-feature-request, and C-enhancement, which
    track discussions on certain features or represent feature proposals.
  \item Feature-gated features: labels prefixed with ``F-,'' which designate
    issues relevant only to language features guarded by feature gates.
\end{itemize}
This filtering process reduced the initial dataset to 320 issues.

Then, we manually investigated each issue to determine whether it is a soundness
issue based on two criteria.
First, we retained an issue only if it contained a Rust code snippet that the
reporter claimed the compiler should reject.
Second, we excluded issues where the buggy behavior was not caused by the
compiler, e.g., those caused by breaking changes in the standard library.
Through this manual inspection, we identified 27 soundness issues.

Finally, we checked whether each issue was a duplicate of another based on
discussions among \rustc{} developers in the issue tracker.
We identified 4 issues as duplicates of others already within our dataset and
subsequently excluded them.
Additionally, 2 issues were duplicates of issues reported before 2022;
we chose to retain these, as their inclusion does not introduce redundancy
within our specific timeframe (see \cref{tab:issues} for these issues and their
original issue numbers).
This resulted in a dataset of 23 unique soundness issues.


After the collection, we compared our dataset with that of Liu et
al.~\cite{liu2025empiricala}, which classifies 22 issues as soundness issues.
Our inspection shows that only 4 issues overlap across the two datasets, while
19 are unique to ours and 18 are unique to the previous study.
Among the 19 issues unique to ours, 4 fall outside the timeframe of the previous
study, 2 were classified under other bug categories, 3 were treated as
duplicates, and 10 were excluded, although we could not find the reasons for
their exclusion in their paper or artifact.
This shows that our study provides a more comprehensive view of soundness
issues, beyond an incremental contribution.
Among the 18 issues unique to the previous study, 8 have type-unrelated area
labels, 1 lacks both C-bug and I-unsound labels, 8 have ``F-'' labels, and 1 is
a duplicate.
We investigated the 9 issues (8 with type-unrelated labels and 1 with no
C-bug/I-unsound label) and found that 7 satisfy our manual inspection criteria
(a Rust code snippet that is claimed the compiler should reject, and buggy
behavior caused by the compiler).
Therefore, we complemented our dataset with these 7 issues, resulting in the
final dataset of 30 issues.
See \cref{tab:issues} for the issues that overlap with or were added from the
previous study.

\section{Soundness Issues in rustc}
\label{sec:bugs}
In this section, we classify the 30 soundness issues according to the affected
feature and the symptom and describe each category with simplified code
snippets.
When describing the issues, we analyze the consequences, triggering features,
and status as well.
We identify five affected features and ten symptoms, as shown in
\cref{tab:features};
we identify four consequences, as shown in \cref{tab:consequences}, and fourteen
triggering features:
{\bf implied} bounds, {\bf assoc}iated types, \C{impl} traits ({\bf IT}s), {\bf
RPIT}s, {\bf tr}ait {\bf obj}ects, higher-rank ({\bf HR}) polymorphism,
associated {\bf const}ants, function pointers ({\bf fn ptr}s), {\bf clo}sures,
{\bf rec}ursive calls, {\bf async}hrony, {\bf try} propagation, {\bf arr}ays,
and {\bf enum}s
(in tables, we abbreviate them as shown in bold).
Note that these affected/triggering features are not intended to systematically
enumerate all features in Rust;
rather, they aid in understanding the soundness issues we collected.
We also classify status into four kinds, as shown in \cref{tab:status-kind}.
Note that an issue is considered fixed if the code given in the issue is
rejected by the latest stable compiler, regardless of its GitHub issue
resolution status.
We use this criterion because an issue may reveal multiple compiler bugs and may
not be closed even after the soundness bug has been fixed due to other remaining
bugs.
\cref{tab:issues} summarizes the characteristics of each issue.
For the sake of brevity, we only present code snippets for some issues;
the omitted ones are available in our supplementary material~\cite{supp}.

\begin{table}[t]
\caption{Classification of affected features and symptoms of soundness issues}
\label{tab:features}
\vspace{-1.15em}
\renewcommand{\arraystretch}{0.9}
  {\scriptsize
\begin{tabular}{l|l}
  Affected feature & Symptoms \\ \hline
  Lifetimes & Missing lifetime checks, Anonymous lifetimes allowed \\
  Traits & Overlapping impls, Missing dyn-compat checks, Orphan impls, Incorrect trait solving \\
  Type inference & Arbitrary type argument selection, Arbitrary hidden type selection \\
  Type well-formedness & Missing well-formedness checks \\
  Type cast & Incorrect type cast allowed \\
\end{tabular}
  }
\vspace{-0.65em}
\end{table}

\begin{table}[t]
\caption{Classification of consequences of soundness issues}
\label{tab:consequences}
\vspace{-1.15em}
\renewcommand{\arraystretch}{0.9}
  {\scriptsize
\begin{tabular}{l|l}
  Consequence & Description \\ \hline
  Memory bug & A memory bug can occur at run time \\
  Breaking change & A benign change in a library can lead to a compilation error in a client code \\
  Inconsistency & A code snippet passes type checking while similar ones are rejected \\
  Misleading behavior & The run-time behavior could be misleading \\
\end{tabular}
  }
\vspace{-0.65em}
\end{table}

\begin{table}[t]
\caption{Classification of status of soundness issues}
\label{tab:status-kind}
\vspace{-1.15em}
\renewcommand{\arraystretch}{0.9}
  {\scriptsize
\begin{tabular}{l|l}
  Status & Description \\ \hline
  Fixed & The compiler has been fixed, and the code snippet is rejected by the latest release (1.93.0) \\
  Confirmed & The C-bug or I-unsound label has been assigned to by a person other than the reporter\\
  Reported & The issue is open but not fixed or confirmed yet \\
  Non-bug & The issue has been closed without being fixed or confirmed \\
\end{tabular}
  }
\vspace{-1.0em}
\end{table}

\begin{table}[t]
\caption{Soundness issues in \rustc{}}
\label{tab:issues}
  \renewcommand{\arraystretch}{0.9}
  \vspace{-1.0em}
  {  \scriptsize
\begin{tabular}{l|l|l|l|l|l}
Issue \# & Affected Feature & Symptom & Consequence & Triggers & Status \\ \hline
\href{https://github.com/rust-lang/rust/issues/96460}{96460}$^\mathsection$& Type inference & Hidden type & Inconsistency & RPIT & Fixed \\
\href{https://github.com/rust-lang/rust/issues/96927}{96927}&Type inference&Type argument&Inconsistency&try&Confirmed$^\mathparagraph$\\
\href{https://github.com/rust-lang/rust/issues/98117}{98117}$^\mathsection$& Lifetimes & Missing lifetime checks & Memory bug & implied, tr obj & Fixed \\
\href{https://github.com/rust-lang/rust/issues/98543}{98543}$^\ddag$&Lifetimes&Missing lifetime checks&Memory bug&implied, assoc&Fixed\\
\href{https://github.com/rust-lang/rust/issues/99158}{99158}$^\ast$&Lifetimes&Missing lifetime checks&Memory bug&implied, fn ptr&Confirmed\\
\href{https://github.com/rust-lang/rust/issues/99554}{99554}&Traits&Orphan&Breaking
change&assoc&Fixed$^{\ast\ast}$\\
\href{https://github.com/rust-lang/rust/issues/100051}{100051}&Lifetimes&Missing lifetime checks&Memory bug&implied, assoc, HR&Reported\\
\href{https://github.com/rust-lang/rust/issues/104005}{104005}$^\mathsection$& Lifetimes & Missing lifetime checks & Memory bug & implied, tr obj & Fixed \\
\href{https://github.com/rust-lang/rust/issues/104763}{104763}$^\mathsection$& Lifetimes & Missing lifetime checks & Inconsistency & const & Fixed \\
\href{https://github.com/rust-lang/rust/issues/104764}{104764}$^\mathsection$& Lifetimes & Missing lifetime checks & Inconsistency & assoc & Fixed \\
\href{https://github.com/rust-lang/rust/issues/105251}{105251}$^\ddag$&Lifetimes&Missing lifetime checks&Inconsistency&assoc, RPIT&Fixed\\
\href{https://github.com/rust-lang/rust/issues/105787}{105787}$^\ddag$&Traits&Overlapping&Inconsistency&assoc, HR&Fixed\\
\href{https://github.com/rust-lang/rust/issues/106040}{106040}&Traits&Incorrect solving&Inconsistency&assoc&Reported\\
\href{https://github.com/rust-lang/rust/issues/107268}{107268}&Type inference&Type argument&Inconsistency&assoc, fn ptr&Fixed\\
\href{https://github.com/rust-lang/rust/issues/108468}{108468}&Lifetimes&Anonymous&Inconsistency&IT, async&Reported\\
\href{https://github.com/rust-lang/rust/issues/109628}{109628}&Lifetimes&Missing lifetime checks&Inconsistency&assoc&Fixed\\
\href{https://github.com/rust-lang/rust/issues/110925}{110925}&Type wf-ness&Missing wf-ness checks&Inconsistency&RPIT&Reported\\
\href{https://github.com/rust-lang/rust/issues/111935}{111935}$^\ddag$&Type inference&Hidden type&Inconsistency&RPIT, rec&Fixed\\
\href{https://github.com/rust-lang/rust/issues/112417}{112417}&Lifetimes&Missing lifetime checks&Memory bug&implied, RPIT&Fixed\\
\href{https://github.com/rust-lang/rust/issues/114061}{114061}&Traits&Overlapping&Inconsistency&assoc, HR&Fixed\\
\href{https://github.com/rust-lang/rust/issues/114389}{114389}$^\dag$&Traits&Overlapping&Memory bug&assoc, tr obj&Confirmed\\
\href{https://github.com/rust-lang/rust/issues/118876}{118876}$^\mathsection$& Lifetimes & Missing lifetime checks & Memory bug & implied, tr obj, HR, clo & Fixed \\
\href{https://github.com/rust-lang/rust/issues/123312}{123312}&Lifetimes&Missing lifetime checks&Inconsistency&clo&Reported\\
\href{https://github.com/rust-lang/rust/issues/126079}{126079}$^\mathsection$& Traits & Missing dyn-compat checks & Memory bug & assoc, tr obj & Fixed \\
\href{https://github.com/rust-lang/rust/issues/133361}{133361}&Traits&Overlapping&Memory bug&assoc, tr obj&Fixed\\
\href{https://github.com/rust-lang/rust/issues/134311}{134311}&Type inference&Hidden type&Inconsistency&assoc, RPIT&Reported\\
\href{https://github.com/rust-lang/rust/issues/136508}{136508}&Type cast&Incorrect cast&Misleading&enum&Reported\\
\href{https://github.com/rust-lang/rust/issues/139406}{139406}&Type inference&Hidden type&Inconsistency&RPIT, rec&Fixed\\
\href{https://github.com/rust-lang/rust/issues/141713}{141713}&Lifetimes&Missing lifetime checks&Memory bug&assoc, HR, clo&Confirmed\\
\href{https://github.com/rust-lang/rust/issues/144045}{144045}&Lifetimes&Missing lifetime checks&Inconsistency&arr&Reported\\
\end{tabular}
  \\[-0.1em] $^\ast$ A duplicate of Issue 25860, which was reported in 2015.
  \\[-0.1em] $^\dag$ A duplicate of Issue 57893, which was reported in 2019.
  \\[-0.1em] $^\ddag$ Overlaps between ours and Liu et al.'s.
  \\[-0.1em] $^\mathsection$ Added after comparison with Liu et al.'s.
  \\[-0.1em] $^\mathparagraph$ This issue was closed as a duplicate of Issue 94902,
  which is not itself a soundness issue but shares the same root cause.
  Since Issue 94902 satisfies the ``confirmed'' criterion, Issue 96927 is
  classified as confirmed.
  \\[-0.6em] $^{\ast\ast}$ A warning occurs instead of an error.
  }
\vspace{-1.5em}
\end{table}

\subsection{Lifetimes}

\subsubsection{Missing Lifetime Checks}

In fourteen soundness issues, the compiler does not check lifetimes properly.
As explained in \cref{sec:background:lifetime}, a reference can be used wherever
its lifetime outlives the expected lifetime, and the compiler must check
lifetimes when references are used to enforce this requirement.
If the compiler fails to check lifetimes, references may be used even after
their lifetimes have ended, incurring memory bugs like use-after-free.
Among the fourteen issues, eight issues actually break memory safety, while six
issues do not result in such misuse of references despite missing lifetime
checks, only exhibiting inconsistent type checking results.
We first discuss the memory-bug-incurring issues and then the others.

{\it Issue 99158.}
In this issue, the compiler does not check implied bounds when coercing function
definitions to function pointers.
In \cref{lst:i99158}, \C{foo} has the implied bound \C{'a: 'static} due to its
second parameter (line~\ref{line:99158:1}).
However, when \C{bar} assigns \C{foo} to \C{f}, this implied bound is not
checked, and the assignment succeeds even though \C{bar} does not have \C{'a:
'static} (line~\ref{line:99158:3}).
By calling \C{f}, \C{bar} can assign \C{'static} to any reference;
if this reference is not valid for the entire execution, dereferencing it would
lead to a memory bug.
Note that explicitly adding the bound \C{'a: 'static} to \C{foo} causes the
assignment to be rejected correctly.

In six other issues, the compiler does not check implied bounds in other
contexts and fails to prevent memory bugs:
it does not check implied bounds when
a trait has a lifetime bound (Issue 98117),
a \emph{self type} (i.e., \C{S} in \C{<S as X>::A}) introduces implied bounds
(Issues 98543 and 100051),
an implementing type in an \C{impl} introduces implied bounds (Issue 104005),
an RPIT hidden type is determined (Issue 112417),
or a closure is coerced (Issue 118876).

{\it Issue 141713.}
In this issue, the compiler does not properly check the lifetime of the associated
type of an HRTB.
In \cref{lst:i141713}, \C{X<T>} declares \C{Fn() -> T} as a \emph{supertrait},
i.e., a type implementing \C{X<T>} must also implement \C{Fn() -> T}
(line~\ref{line:i141713:1}).
\C{Fn() -> T} is a built-in trait automatically implemented by every closure
that takes no parameters and returns \C{T}.
It has an associated type named \C{Output}, which is set to \C{T}.
Additionally, every type implementing \C{Fn() -> T} also implements \C{X<T>}
(line~\ref{line:i141713:2}).
In \C{foo}, \C{F} has an HRTB, but it is unclear what \C{F::Output} denotes;
the compiler allows using \C{F::Output} as a reference with \C{'static} in
\C{foo} (line~\ref{line:i141713:3}), while it allows using a reference with any
lifetime as \C{F::Output} in \C{bar} (line~\ref{line:i141713:4}).
Note that \C{|| \&0} is a closure that takes no parameters and returns \C{\&0}.

{\it Issue 144045.}
In this issue, the compiler does not check the lifetime of a length-0 array.
In \cref{lst:i144045}, \C{foo} returns \C{[v; 1]}, which is a length-1 array
with \C{v} as its element, and it is rejected because \C{'a} does not outlive
\C{'static} (line~\ref{line:144045:1}).
On the other hand, \C{bar} is accepted, even though the only difference from
\C{foo} is that it returns \C{[v; 0]}, which is a length-0 array
(line~\ref{line:144045:2}).
This issue does not cause memory bugs because \C{bar} does not actually return
any reference.
Nevertheless, this behavior might seem surprising, as developers usually expect
the type checking of array types to be independent of their lengths.

Similarly, five other issues also reveal cases where the compiler does not check
lifetimes but still prevents memory bugs.
In particular, it does not check lifetimes when
an associated constant or type is accessed (Issues 104763 and 104764),
an RPIT has nested \C{impl} traits (Issue 105251),
a type parameter's associated type has a trait bound (Issue 109628),
or a struct value is constructed inside a closure (Issue 123312).

\begin{figure}[t]
\begin{lstlisting}[caption={Issue 99158}, label={lst:i99158}]
fn foo<'a>(v: &'a i32, u: &'static &'a i32) -> &'static i32 { v } $\label{line:99158:1}$
fn bar<'a>(v: &'a i32) -> &'static i32 { $\label{line:99158:2}$
    let f: fn(&'a i32, _) -> &'static i32 = foo; f(v, &&0) } $\label{line:99158:3}$
\end{lstlisting}
\vspace{-1.2em}
\begin{lstlisting}[caption={Issue 141713}, label={lst:i141713}]
trait X<T>: Fn() -> T {} $\label{line:i141713:1}$
impl<T, F: Fn() -> T> X<T> for F {} $\label{line:i141713:2}$
fn foo<F>(v: F, u: F::Output) -> &'static i32 where for<'a> F: X<&'a i32>{u}$\label{line:i141713:3}$
fn bar<'a>(v: &'a i32) -> &'static i32 { foo(|| &0, v) } $\label{line:i141713:4}$
\end{lstlisting}
\vspace{-1.2em}
\begin{lstlisting}[caption={Issue 144045}, label={lst:i144045}]
fn foo(v: &'a i32) -> [&'static i32; 1] { [v; 1] } // error $\label{line:144045:1}$
fn bar(v: &'a i32) -> [&'static i32; 0] { [v; 0] } // ok $\label{line:144045:2}$
\end{lstlisting}
\vspace{-1.2em}
\begin{lstlisting}[caption={Issue 108468}, label={lst:i108468}]
trait X<T> {} $\label{line:i108468:1}$
fn foo(v: impl X<&i32>) {} // error $\label{line:i108468:2}$
fn bar<'a>(v: impl X<&'a i32>) {} // ok $\label{line:i108468:3}$
async fn baz(v: impl X<&i32>) {} // ok $\label{line:i108468:4}$
\end{lstlisting}
\vspace{-1.2em}
\end{figure}

\subsubsection{Anonymous Lifetimes Allowed}

In one soundness issue, the compiler unexpectedly allows \emph{anonymous
lifetimes}.
In Rust, when the compiler can automatically introduce necessary lifetime
parameters according to predefined rules~\cite{rust-lifetime-elision},
programmers can use anonymous lifetimes.
For example, \C{foo} and \C{bar} are equivalent in the following code:

\vspace{-0.5em}
\begin{lstlisting}
fn foo(v: &i32) {} fn bar<'a>(v: &'a i32) {}
\end{lstlisting}
\vspace{-0.5em}

\noindent
However, anonymous lifetimes are explicitly disallowed in some contexts, and if
the compiler allows them, it may be considered a soundness bug.

{\it Issue 108468.}
In this issue, using an \C{async} function causes the compiler to allow
anonymous lifetimes.
In \cref{lst:i108468}, the function \C{foo} is rejected because anonymous
lifetimes must not occur in \C{impl} traits (line~\ref{line:i108468:2}), while
\C{bar} is accepted (line~\ref{line:i108468:3}).
However, the \C{async} function \C{baz} passes type checking even though it
contains an anonymous lifetime in an \C{impl} trait (line~\ref{line:i108468:4}).
%
%

\subsection{Traits}

\subsubsection{Overlapping Implementations}

In four soundness issues, the compiler allows \emph{overlapping impls}, i.e.,
those implementing the same trait for the same type.
Rust prohibits overlapping \C{impl}s because they create ambiguity when deciding
which \C{impl} to use during method resolution.
For example, the following code is rejected:

\vspace{-0.5em}
\begin{lstlisting}
trait X { fn foo(self) -> i32; } struct S;
impl X for S{fn foo(self)->i32{0}}impl X for S{fn foo(self)->i32{1}} // error
\end{lstlisting}
\vspace{-0.5em}

\noindent
Therefore, if code with overlapping \C{impl}s passes type checking, it can be
considered a soundness bug.
Among the four issues in this category, two lead to memory bugs, while two
only cause inconsistency.

{\it Issue 114389.}
In this issue, the compiler allows a user-defined \C{impl} to overlap with a
\emph{built-in impl}---for every trait \C{X}, the type \C{dyn X} automatically
implements \C{X}, called a built-in \C{impl}.
In \cref{lst:i114389}, every type implements \C{X}, even including dynamically
sized types due to the bound \C{?Sized} on \C{T} (line~\ref{line:114389:2}).
In \C{foo}, since \C{<T as X>::A} and \C{\&'static i32} are the same type, it
can return the given value as is (line~\ref{line:114389:3}).
On the other hand, when \C{bar} calls \C{foo}, the type argument is \C{dyn X<A =
\&'a i32>}, so the parameter type \C{<T as X>::A} becomes \C{\&'a i32}.
Consequently, \C{bar} can use a value of \C{\&'a i32} as \C{\&'static i32} by
passing it to \C{foo} (line~\ref{line:114389:4}).
This issue arises because \C{dyn X<A = \&'a i32>} has two overlapping \C{impl}s
of \C{X}:
the one with \C{A = \&'static i32} defined in line~\ref{line:114389:2} and the
built-in \C{impl} with \C{A = \&'a i32}.

In another issue (Issue 133361), two user-defined \C{impl}s for trait objects
overlap and similarly incur memory bugs by setting the same associated type to
two different types.
In two other issues, the compiler fails to detect overlapping \C{impl}s when
higher-rank polymorphism is involved, but memory bugs do not occur
because the compiler always selects one \C{impl} and ignores the other (Issue
105787),
or when the associated type is used, the compiler recognizes that two different
types are possible and rejects the code (Issue 114061).

\begin{figure}[t]
\begin{lstlisting}[caption={Issue 114389}, label={lst:i114389}]
trait X { type A; } impl<T: ?Sized> X for T { type A = &'static i32; } $\label{line:114389:2}$
fn foo<T: ?Sized>(v: <T as X>::A) -> &'static i32 { v } $\label{line:114389:3}$
fn bar<'a>(v: &'a i32) -> &'static i32 { foo::<dyn X<A = &'a i32>>(v) } $\label{line:114389:4}$
\end{lstlisting}
\vspace{-2em}
\end{figure}

\subsubsection{Missing Dyn-compatibility Checks}

In one soundness issue (Issue 126079), the compiler does not check the
\emph{dyn-compatibility} of a trait.
To use a type \C{dyn T}, the trait \C{T} must satisfy dyn-compatibility, which
requires several conditions, such as not having a method return type containing
\C{Self}.
However, the compiler does not check dyn-compatibility when an associated type
of a supertrait, referred to using \C{Self}, is used as a return type.
This allows a single return type to be interpreted as two different types,
leading to memory bugs.

\subsubsection{Orphan Implementations}

In one soundness issue, the compiler allows \emph{orphan impls}, i.e., those
implementing external traits for external types.
In Rust, a \emph{crate} is a compilation unit, and items defined in other crates
are considered external.
If orphan \C{impl}s are allowed, a seemingly innocent change in a library can
prevent a dependent program from compiling:

\vspace{-0.5em}
\begin{lstlisting}
// crate a $\label{line:orphan:1}$
trait X {} struct S; $\label{line:orphan:2}$
// crate b (depends on crate a) $\label{line:orphan:3}$
trait Y {} struct R; $\label{line:orphan:4}$
impl a::X for a::S {} // error $\label{line:orphan:5}$
impl a::X for R {} impl Y for a::S {} // ok $\label{line:orphan:6}$
\end{lstlisting}
\vspace{-0.5em}

\noindent
While the crate \C{a} currently does not implement \C{X} for \C{S}, it can
freely add an \C{impl} in the future.
If the crate \C{b} is allowed to implement \C{X} for \C{S}, such an update in
\C{a} becomes an abrupt breaking change, making \C{b} uncompilable due to two
overlapping \C{impl}s of \C{X} for \C{S}.
To prevent this in advance, the orphan implementation is rejected
(line~\ref{line:orphan:5}), while implementing \C{X} for its own type \C{R} or
its own trait \C{Y} for \C{S} is accepted (line~\ref{line:orphan:6}).
Therefore, the issue in this category leads to a failure in preventing future
breaking changes.

{\it Issue 99554.}
In this issue, the compiler allows an orphan \C{impl} when an associated type is
used.
In \cref{lst:i99554}, \C{a} defines \C{X} (line~\ref{line:i99554:2}).
In \C{b}, implementing \C{X} for any type \C{T} that implements \C{Y} is
considered orphan and correctly rejected, as such \C{T} can be external
(line~\ref{line:i99554:5}).
However, writing \C{<T as Y>::A}, which is equivalent to \C{T}, causes the
compiler to accept the \C{impl} (line~\ref{line:i99554:6}).

\subsubsection{Incorrect Trait Solving}

In one soundness issue, the compiler draws an incorrect conclusion in
\emph{trait solving}, i.e., deciding whether a given type implements a certain
trait.

{\it Issue 106040.}
In this issue, trait solving succeeds despite the existence of cyclic
requirements when associated types have trait bounds.
In \cref{lst:i106040}, the compiler needs to determine whether \C{S} implements
\C{Unpin} (line~\ref{line:i106040:4}), where \C{Unpin} is a trait defined in the
standard library.
This requires \C{S: X<A = S>} (line~\ref{line:i106040:3}), which in turn
requires \C{S: Unpin} (line~\ref{line:i106040:2}), forming a cycle.
However, the compiler concludes that \C{S} implements \C{Unpin} and accepts the
entire program.
This result can be considered inconsistent because the compiler successfully
detects the cycle and rejects the code when the bound is directly applied to the
type parameter, instead of the associated type, as follows:

\vspace{-0.5em}
\begin{lstlisting}[firstnumber=2]
impl<T: Unpin> X for T { type A = T; }
\end{lstlisting}
\vspace{-0.5em}

\begin{figure}[t]
\begin{lstlisting}[caption={Issue 99554}, label={lst:i99554}]
// crate a $\label{line:i99554:1}$
trait X<T, U> {} $\label{line:i99554:2}$
// crate b $\label{line:i99554:3}$
struct S; trait Y { type A; } impl<T> Y for T { type A = T; } $\label{line:i99554:4}$
impl<T: Y> a::X<S, T> for T {} // error $\label{line:i99554:5}$
impl<T: Y> a::X<S, T> for <T as Y>::A {} // ok $\label{line:i99554:6}$
\end{lstlisting}
\vspace{-1.2em}
\begin{lstlisting}[caption={Issue 106040}, label={lst:i106040}]
trait X { type A where Self: Unpin; } $\label{line:i106040:1}$
impl<T> X for T { type A = T where T: Unpin; } $\label{line:i106040:2}$
struct S; impl Unpin for S where S: X<A = S> {} $\label{line:i106040:3}$
fn foo<T: Unpin>() {} fn bar() { foo::<S>() } $\label{line:i106040:4}$
\end{lstlisting}
\vspace{-1.2em}
\begin{lstlisting}[caption={Issue 96927}, label={lst:i96927}]
fn foo<T>() -> Option<T> { None } $\label{line:i96927:1}$
fn bar() -> Option<i32> { foo(); None } // error $\label{line:i96927:2}$
fn baz() -> Option<i32> { foo()?; None } // ok $\label{line:i96927:3}$
\end{lstlisting}
\vspace{-2em}
\end{figure}

\subsection{Type Inference}

\subsubsection{Arbitrary Type Argument Selection}

In two soundness issues, the compiler arbitrarily chooses a type argument during
type inference even when multiple types are possible.
In Rust, type inference of a type argument should succeed only when the possible
type is unique.
In the following example, \C{foo(0)} is accepted because \C{i32} is the only
possible type, while \C{baz()} is rejected because any type would be valid:

\vspace{-0.5em}
\begin{lstlisting}
fn foo<T>(t: T) {} fn bar() { foo(0); } // ok
fn baz<T>() {} fn qux() { baz(); } // error
\end{lstlisting}
\vspace{-0.5em}

\noindent
If the compiler arbitrarily chooses one among multiple possible types, it can be
considered inconsistent, and the run-time behavior may deviate from the
developer's expectation.

{\it Issue 96927.}
In this issue, type inference selects an arbitrary type when the try propagation
operator (\C{?}) is used.
In \cref{lst:i96927}, \C{foo} returns \C{Option<T>}, which expresses the
optional existence of a value, allowing \C{Some(v)} for a value or \C{None} for
no value (line~\ref{line:i96927:1}).
Since \C{bar} does not use the return value, the type argument can be any type,
and type inference fails (line~\ref{line:i96927:2}).
On the other hand, \C{baz} uses \C{?}, which immediately returns \C{None} if the
result is \C{None} and extracts the inner value otherwise
(line~\ref{line:i96927:3}).
The compiler still has no information about the type, but chooses an arbitrary
one---\C{!} (signifying that the function never returns) in recent versions, or
\C{()} (the empty tuple) in older versions---and accepts the code.

In another issue (Issue 107268), the compiler chooses an arbitrary type as a
type argument when a function with a lifetime parameter is coerced to a function
pointer.

\subsubsection{Arbitrary Hidden Type Selection}

In four soundness issues, type inference chooses an arbitrary type as the
hidden type for an RPIT even when multiple types are possible.
Such issues are also considered inconsistent and may lead to unexpected run-time
behavior.

{\it Issue 139406.}
In this issue, type inference selects an arbitrary hidden type when recursive
calls exist.
The function \C{foo} passes the result of a recursive call to another recursive
call (line~\ref{line:i139406:3}).
While both \C{T} and \C{!} are valid choices, the compiler chooses \C{T} and
accepts the code.

In three other issues, the hidden type is arbitrarily selected when
a built-in type \C{PhantomData} (Issue 96460),
recursive calls (Issue 111935), or associated types (Issue 134311) are involved.

\begin{figure}[t]
\begin{lstlisting}[caption={Issue 139406}, label={lst:i139406}]
trait X {} impl<T> X for T {} $\label{line:i139406:1}$
fn foo<T>(v: T) -> impl X { foo::<T>(foo::<T>(v)) } $\label{line:i139406:3}$
\end{lstlisting}
\vspace{-1em}
\begin{lstlisting}[caption={Issue 110925}, label={lst:i110925}]
trait X<T> {} impl<T> X<T> for i32 {} trait Y {} struct S<T: Y>(T); $\label{line:i110925:1}$
fn foo() -> impl X<S<i32>> { 0 } // ok $\label{line:i110925:2}$
fn bar(v: impl X<S<i32>>) {} // error $\label{line:i110925:3}$
\end{lstlisting}
\vspace{-1em}
\begin{lstlisting}[caption={Issue 136508}, label={lst:i136508}]
enum E { A } enum F { B(i32) } $\label{line:i136508:1}$
fn foo() { let x = E::A as i32; // ok $\label{line:i136508:2}$
           let y = F::B(0) as i32; // error $\label{line:i136508:3}$
           let z = F::B as i32; } // ok $\label{line:i136508:4}$
\end{lstlisting}
\vspace{-1em}
\end{figure}

\subsection{Type Well-formedness}

{\it Issue 110925.}
In this issue, the compiler does not check the well-formedness of a type in an
RPIT.
In \cref{lst:i110925}, \C{S<i32>} is not well-formed because no type implements
\C{Y}, while \C{S<T>} requires \C{T: Y} (line~\ref{line:i110925:1}).
However, \C{foo} is accepted despite using \C{S<i32>} in the RPIT
(line~\ref{line:i110925:2}).
The compiler rejects the use of \C{S<i32>} in other contexts, including APITs
(line~\ref{line:i110925:3}).
This inconsistency does not lead to memory bugs because a value of \C{S<i32>}
cannot be constructed.

\subsection{Type Cast}

{\it Issue 136508.}
This issue allows casting an enum constructor to an integer.
In Rust, an enum consists of multiple \emph{variants}, each of which may have
associated data.
In \cref{lst:i136508}, \C{E} has one variant \C{A} without data, and \C{F} has
one variant \C{B} with an \C{i32} value (line~\ref{line:i136508:1}).
Rust allows casting \C{E::A}, which has no data, to an integer
(line~\ref{line:i136508:2}), but disallows casting \C{F::B(0)}, which contains
data (line~\ref{line:i136508:3}).
Somewhat surprisingly, casting \C{F::B} to an integer is accepted
(line~\ref{line:i136508:4}).
In fact, this behavior is intended according to the Rust Reference, as \C{F::B}
is an enum constructor and treated as a function;
casting it to an integer gives its address.
However, the reporter of this issue claims that this behavior is misleading
because a developer might attempt to cast \C{F::B} under the mistaken impression
that it is a variant without data.

\section{Findings from Soundness Issues}
\label{sec:findings}
In this section, we discuss findings from the collected soundness issues by
answering the following research questions:
\begin{itemize}[leftmargin=*]
  \item {\bf RQ1. Symptoms and Consequences:} What are the symptoms and
    consequences of soundness issues?
  \item {\bf RQ2. Triggering Features:} Which language features trigger
    soundness issues?
  \item {\bf RQ3. Bug Consensus:} For which soundness issues is there consensus
    that they are bugs?
  \item {\bf RQ4. Issue Lifecycles:} How long does it take to discover and fix
    soundness issues?
\end{itemize}

\subsection{RQ1. Symptoms and Consequences}

\cref{tab:symptoms-consequences} shows the number of soundness issues per
symptom and affected feature (the two rightmost columns), while also displaying
how often each symptom leads to each consequence (the four middle columns).
Lifetimes are the most frequently affected feature (15 issues), followed by
traits (7 issues) and type inference (6 issues).
In terms of the symptoms, missing lifetime checks are the most common (14
issues), followed by overlapping \C{impl}s (4 issues) and arbitrary hidden type
selection (4 issues).
The issues mainly lead to inconsistencies (17 issues) and memory bugs (11
issues).

{\bf Finding 1. The Rust compiler has soundness issues that lead to memory bugs,
caused by missing lifetime checks or overlapping \C{impl}s.}
Although not all soundness issues result in memory bugs, 11 of them do.
Among these, 8 are caused by missing lifetime checks, which is expected as
lifetimes are key to enforcing the valid use of pointers.
On the other hand, 2 are caused by overlapping \C{impl}s, and 1 is caused by
missing dyn-compatibility checks.
The primary role of traits is to enable polymorphism, which is not directly
relevant to pointer usage.
However, overlapping \C{impl}s and missing dyn-compatibility checks may allow a
single type to be interpreted as different types, and such type confusion can
result in memory bugs.
This shows that the research community should recognize the importance of the
trait system in ensuring memory safety, whereas previous
studies~\cite{jung2017rustbelt, wagner2025linearity} have mainly focused on
formalizing the ownership and borrowing system, including lifetimes.

{\bf Finding 2. Rust programmers often believe that the compiler should ensure
consistency by rejecting even programs that do not incur memory bugs.}
For example, type inference choosing an arbitrary type cannot cause memory bugs.
Nevertheless, many developers expect the compiler to reject such cases,
following the consistent rule that inference should succeed only when a unique
valid type exists.
Similarly, in Issue 144045, although empty arrays cannot carry any pointers,
which is a valid reason for lifetime checks to be exempted, developers may
expect them to be treated in a way consistent with non-empty arrays.

\subsection{RQ2. Triggering Features}

\newcommand{\zero}{{\color{lightgray}0}}
\begin{table}[t]
\caption{Symptoms and consequences of soundness issues}
  \label{tab:symptoms-consequences}
\vspace{-1.1em}
  \renewcommand{\arraystretch}{0.99}
{\scriptsize
\begin{tabular}{l|l|l|r|r|r|r|r|r}
  \multicolumn{3}{c|}{} & \multicolumn{4}{c|}{\bf Consequence} &
  \multicolumn{2}{c}{\multirow{2}{*}{\bf Total}} \\
  \cline{4-7}
  \multicolumn{3}{c|}{} & Memory & Breaking & Inconsistency & Misleading & \multicolumn{2}{c}{} \\
  \hline
  \multirow{9}{*}{\bf Symptom}
  &\multirow{2}{*}{Lifetimes}
  & Missing lifetime checks & 8 & \zero & 6 & \zero & \bf 14 & \multirow{2}{*}{\bf 15} \\
  && Anonymous lifetimes & \zero & \zero & 1 & \zero & \bf 1 \\
  \cline{2-3} \cline{8-9}
  &\multirow{4}{*}{Traits}
  & Overlapping {\tt impl}s & 2 & \zero & 2 & \zero & \bf 4 & \multirow{4}{*}{\bf 7} \\
  && Missing dyn-cmpt checks & 1 & \zero & \zero & \zero & \bf 1 \\
  && Orphan {\tt impl}s & \zero & 1 & \zero & \zero & \bf 1 \\
  && Incorrect trait solving & \zero & \zero & 1 & \zero & \bf 1 \\
  \cline{2-3} \cline{8-9}
  &\multirow{2}{*}{Inference}
  & Type argument & \zero & \zero & 2 & \zero & \bf 2 & \multirow{2}{*}{\bf 6} \\
  && Hidden type & \zero & \zero & 4 & \zero & \bf 4 \\
  \cline{2-3} \cline{8-9}
  &\multirow{1}{*}{Wf-ness}
  & Missing wf-ness checks & \zero & \zero & 1 & \zero & \bf 1 & \multirow{1}{*}{\bf 1} \\
  \cline{2-3} \cline{8-9}
  &\multirow{1}{*}{Cast}
  & Incorrect cast & \zero & \zero & \zero & 1 & \bf 1 & \multirow{1}{*}{\bf 1} \\
  \hline
  \multicolumn{3}{c|}{\bf Total} & \bf 11 & \bf 1 & \bf 17 & \bf 1 & \multicolumn{2}{c}{\bf 30} \\
\end{tabular}
}
\vspace{-0.6em}
\end{table}

\begin{table}[t]
\caption{Triggering features of soundness issues}
\label{tab:triggers}
\vspace{-1.1em}
  \renewcommand{\arraystretch}{0.99}
{\scriptsize
\begin{tabular}{l|l|@{}r@{\,}|@{\,}r@{\,}|@{\,}r@{\,}|@{\,}r@{\,}|@{\,}r@{\,}|@{\,}r@{\,}|@{\,}r@{\,}|@{\,}r@{\,}|@{\,}r@{\,}|@{\,}r@{\,}|@{\,}r@{\,}|@{\,}r@{\,}|@{\,}r@{\,}|@{\,}r@{\,}|r}
  \multicolumn{2}{c|@{}}{} & \multicolumn{14}{c|}{\bf Triggering Feature} &
  \multirow{2}{*}{\bf Total} \\
  \cline{3-16}
  \multicolumn{2}{c|@{}}{} & \,implied & assoc & IT & RPIT & tr obj & HR & const & fn ptr & clo & rec & async & try & arr & enum & \\
  \hline
  \multirow{5}{*}{\bf Affected}
  & Lifetimes & 7 & 6 & 1 & 2 & 3 & 3 & 1 & 1 & 3 & \zero & 1 & \zero & 1 & \zero & \bf 15 \\
  & Traits & \zero & 7 & \zero & \zero & 3 & 2 & \zero & \zero & \zero & \zero & \zero & \zero & \zero & \zero & \bf 7 \\
  & Inference & \zero & 2 & \zero & 4 & \zero & \zero & \zero & 1 & \zero & 2 & \zero & 1 & \zero & \zero & \bf 6 \\
  & Wf-ness & \zero & \zero & \zero & 1 & \zero & \zero & \zero & \zero & \zero & \zero & \zero & \zero & \zero & \zero & \bf 1 \\
  & Cast & \zero & \zero & \zero & \zero & \zero & \zero & \zero & \zero & \zero & \zero & \zero & \zero & \zero & 1 & \bf 1 \\
  \hline
  \multirow{4}{*}{\bf Consequence}
  & Memory & 7 & 6 & \zero & 1 & 6 & 3 & \zero & 1 & 2 & \zero & \zero & \zero & \zero & \zero & \bf 11 \\
  & Breaking & \zero & 1 & \zero & \zero & \zero & \zero & \zero & \zero & \zero & \zero & \zero & \zero & \zero & \zero & \bf 1 \\
  & Inconsistency & \zero & 8 & 1 & 6 & \zero & 2 & 1 & 1 & 1 & 2 & 1 & 1 & 1 & \zero & \bf 17 \\
  & Misleading & \zero & \zero & \zero & \zero & \zero & \zero & \zero & \zero & \zero & \zero & \zero & \zero & \zero & 1 & \bf 1 \\
  \hline
  \multicolumn{2}{c|@{}}{\bf Total} & \bf 7 & \bf 15 & \bf 1 & \bf 7 & \bf 6 & \bf 5 & \bf 1 & \bf 2 & \bf 3 & \bf 2 & \bf 1 & \bf 1 & \bf 1 & \bf 1 & \bf 30 \\
\end{tabular}
}
\vspace{-0.6em}
\end{table}

\begin{table}[t]
\caption{Status of soundness issues}
\label{tab:status}
\vspace{-1.1em}
  \renewcommand{\arraystretch}{0.99}
{\scriptsize
\begin{tabular}{l|l|r|r|r|r|r}
  \multicolumn{2}{c|}{} & \multicolumn{4}{c|}{\bf Status} & \multirow{2}{*}{\bf Total} \\
  \cline{3-6}
  \multicolumn{2}{c|}{} & Fixed & Confirmed & Reported & Non-bug & \\
  \hline
  \multirow{4}{*}{\bf Consequence}
  & Memory & 7 & 3 & 1 & \zero & \bf 11 \\
  & Breaking & 1 & \zero & \zero & \zero & \bf 1 \\
  & Inconsistency & 10 & 1 & 6 & \zero & \bf 17 \\
  & Misleading & \zero & \zero & 1 & \zero & \bf 1 \\
  \hline
  \multicolumn{2}{c|}{\bf Total} & \bf 18 & \bf 4 & \bf 8 & \bf 0 & \bf 30 \\
\end{tabular}
}
\vspace{-0.6em}
\end{table}

\cref{tab:triggers} shows the number of issues that require each feature as a
trigger, grouped by affected features (the upper five rows) and consequences
(the lower four rows).
Note that the rightmost column does not show the sum of the numbers in the
middle columns;
it shows the number of issues in each row, which is identical to the totals in
\cref{tab:symptoms-consequences}.
Associated types are the most frequently involved feature (15 issues), followed
by implied bounds (7 issues), RPITs (7 issues), and higher-rank polymorphism (5
issues).

{\bf Finding 3. Compiler developers should carefully handle associated types
wherever they can occur.}
Half of the issues require associated types as a trigger.
However, it is relatively rare that associated types alone cause soundness
issues;
rather, they often become a way to exploit edge cases when using other features.
For example, in Issue 99554, using an associated type instead of the type itself
causes the compiler to fail to detect an orphan \C{impl}.
Therefore, developers should consider the possibility of using associated types
with caution.

{\bf Finding 4. The interaction between orthogonal features, especially
lifetimes and traits, makes it difficult for type checking to be sound.}
Most issues that affect lifetimes require trait-related features as triggers:
associated types in 6 issues, \C{impl} traits in 1 issue, and RPITs in 2 issues.
This shows that correctly implementing lifetime checks becomes especially
challenging when traits are involved.
Therefore, future work is encouraged to study lifetimes and traits in a unified
manner, instead of treating them separately.

{\bf Finding 5. Correct treatment of implied bounds and trait objects is
extremely important in ensuring memory safety.}
Implied bounds are involved in 7 issues, all of which lead to memory bugs.
Implied bounds are automatically introduced by the compiler;
to retain memory safety, \rustc{} developers should carefully validate that all
the implied bounds introduced at the definition site are also checked at the use
site.
As existing formalizations of the borrowing system~\cite{jung2017rustbelt,
wagner2025linearity} do not include implied bounds, this result suggests that
they should be considered in future research.
Similarly, trait objects trigger 6 issues, all of which lead to memory bugs.
As trait objects automatically introduce built-in \C{impl}s, they make the
detection of overlapping \C{impl}s even more challenging.
Although Milewski~\cite{milewski2015traits}'s formalization of the trait system
includes coherence rules, the rules do not cover trait objects, necessitating
future work to address this gap.

Note that we do not claim that soundness issues involving specific features are
necessarily disproportionate.
Lifetimes and traits are pervasively used in real-world Rust programs, which may
be one reason why they are prevalently involved in soundness issues.
We believe this does not undermine, but rather emphasizes, the importance of our
findings;
if they are frequently used, related soundness issues are especially critical
and require more attention.

These findings are mostly specific to Rust, rather than applicable to
programming languages in general, due to Rust's unique features.
Notably, lifetimes have no equivalents in other mainstream languages, and traits
resemble type classes in Haskell but have no exact equivalents in
object-oriented languages.
For example, bug categories from a study on compiler bugs in JVM-targeting
languages~\cite{chaliasos2021jvm} include incorrect inference and incorrect
coercion, but nothing corresponding to lifetimes or traits.

\subsection{RQ3. Bug Consensus}

\cref{tab:status} shows the number of issues with a certain status per
consequence.
Overall, 18 have been fixed, 4 have been confirmed, 8 have been reported, and
none are non-bugs.
Among 12 issues that lead to memory bugs or breaking changes, 11 have been fixed
or confirmed, indicating that they are real bugs.
This is not surprising because memory safety is the main goal of Rust, and
breaking changes are caused by a clear violation of orphan rules.

{\bf Finding 6. The \rustc{} developers are inclusive in discussing soundness
issues even when they do not lead to memory bugs.}
Among 18 issues that lead to inconsistencies and misleading behavior, only 11
have been fixed or confirmed.
Nevertheless, none of the remaining issues have been classified as non-bugs.
This may reveal the community's inclusive nature in judging soundness bugs.
A notable example is Issue 136508, where an enum constructor can be cast to an
integer.
Even though the Rust Reference explicitly describes this behavior and one
developer clarified this point in the discussion, the issue remains open, and
another developer even added the T-lang, T-compiler, and A-coercions labels.

\subsection{RQ4. Issue Lifecycles}

\begin{table}[t]
\caption{Statistics regarding the lifecycles of the soundness issues (in days)}
\label{tab:duration}
\vspace{-1.0em}
  \renewcommand{\arraystretch}{0.99}
  {\scriptsize
\begin{tabular}{l|rr|r}
              & \multicolumn{2}{c|}{\textbf{All}} & \multicolumn{1}{c}{\textbf{Fixed}}
\\
 \cline{2-4}
\topspace
\textbf{}     & Testable-Introduced           & Introduced-Discovery          & Discovery-Fixed       \\
\hline
Minimum           & 0                        & 1                       & 2                  \\
Maximum           & 2,813                     & 3,552                    & 855                \\
Median        & 855                        & 1,079                    & 345                \\
Average       & 866                    & 1,291                  & 324             
\end{tabular}
  }
\vspace{-1.0em}
\end{table}

\begin{figure}[t]
\centering
\includegraphics[width=0.99\textwidth]{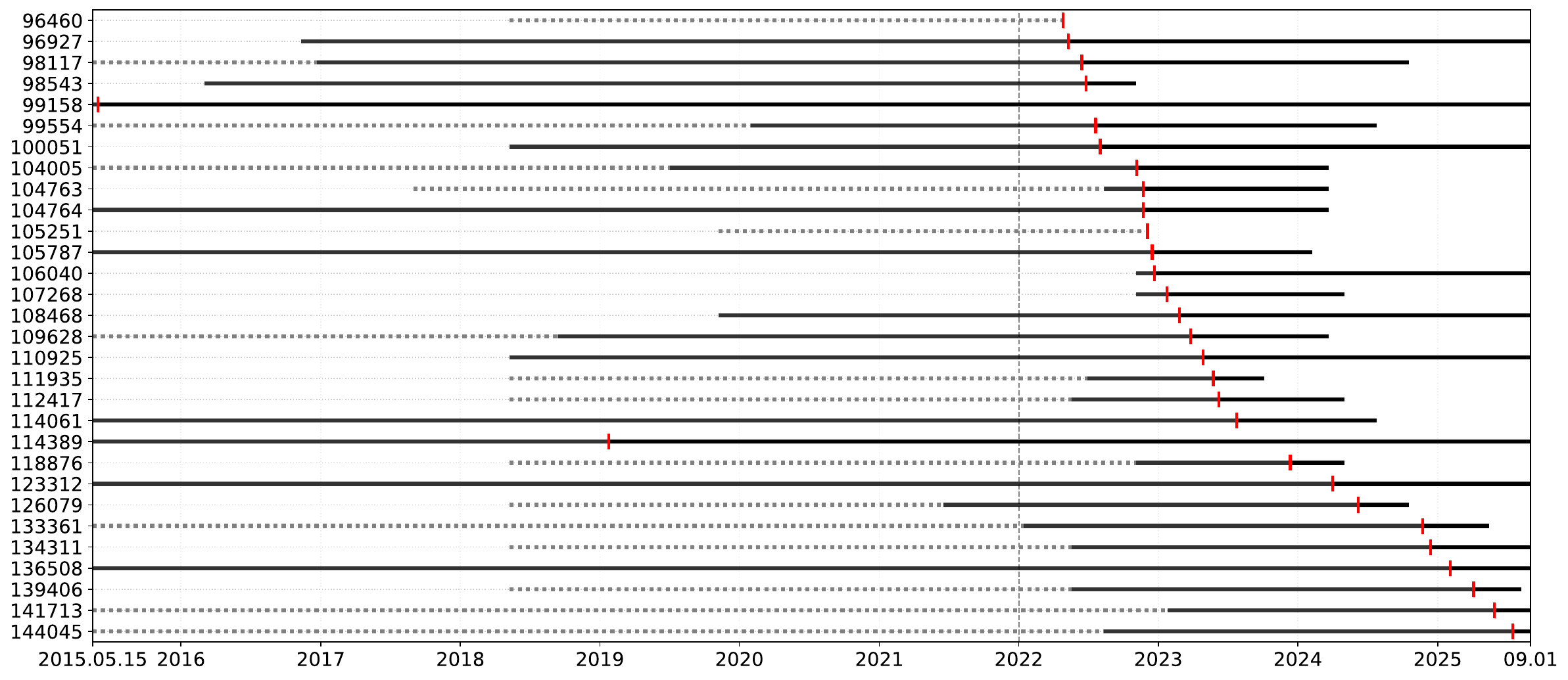}
\vspace{-1.0em}
\caption{Lifecycles of the soundness issues. Each row shows one issue: gray dashed lines indicate the period before introduction, solid black lines indicate the period from introduction to fix or study end, red markers indicate discovery dates, and the vertical dashed line marks the start of issue collection.}
\vspace{-1.0em}
\label{fig:timing}
\end{figure}

We analyze the lifecycle of each issue by identifying stable releases of
\rustc{} where it is testable, introduced, or fixed.
The earliest testable version is the one where the immediately preceding version
rejects the code with a message that a feature is unstable;
the introduction version is the earliest stable release in which the issue can
be triggered.
We tested each version of \rustc{} against each code snippet to find these
versions.
In addition, we determine the discovery date as the issue's creation date.
A few exceptions to these criteria are:
(1) Issues 96460 and 105251 are not reproducible in stable releases, so we identify the
nightly versions where the issue occurs;
(2) Issues 99158 and 114389 are duplicates of 25860 and 57893, respectively, so
we refer to the earlier ones to avoid considering their discovery dates absurdly
late.

%
%
%

\cref{tab:duration} summarizes the statistics regarding the lifecycles of the
soundness issues.
The two middle columns present the statistics of the durations (in days) from
the earliest testable version to introduction and from introduction to discovery
for all 30 issues.
The rightmost column reports the statistics for the durations from discovery to
fix for the 18 fixed issues.

\cref{fig:timing} visualizes the lifecycles of the soundness issues.
The x-axis represents time, starting from the release of \rustc{} 1.0.0,
with the vertical dashed line at 2022 marking the beginning of our issue
collection period.
For each issue arranged along the y-axis, gray dashed lines span the duration
from the earliest testable version to the introduction,
and solid black lines span the duration from the introduction to the fix or the
end of the collection period.
Finally, red vertical markers denote the discovery dates.

{\bf Finding 7. Soundness issues are typically discovered long after they are
introduced, persisting across multiple releases.}
The discovery of these issues took an average of 1,291 days, with a maximum of
3,552 days.
Soundness issues are particularly difficult to identify because they often
require peculiar combinations of multiple language features that developers
rarely exercise in practice, and allow compilation to succeed silently, not
necessarily accompanied by immediately observable problems.
These results indicate a critical need for systematic testing techniques to
detect soundness issues in \rustc{}.

%

{\bf Finding 8. New features are often stabilized before undergoing sufficient
testing.}
Seven issues (96927, 98543, 100051, 106040, 107268, 108468, and 110925) were
introduced simultaneously with the stabilization of the features required to
trigger them.
Before stabilizing new features, compiler developers must ensure the soundness
of the type checker, specifically by investigating potential interactions
between the new feature and existing language constructs.

We believe Findings 7 and 8 are general phenomena across different programming
languages, and our contribution is to empirically confirm that they also apply
to Rust.
These findings are especially meaningful given Rust's emphasis on type
soundness.

%

\section{Potential Oracles for Testing Type Soundness}
\label{sec:oracles}
\subsection{Implementations}

\subsubsection{AddressSanitizer and Miri}

AddressSanitizer~\cite{serebryany2012addresssanitizer} is a memory bug detector
based on instrumentation, and
Miri~\cite{jung2026miri, gh-miri} is a Rust interpreter designed to detect
undefined behavior, including memory bugs.
They can serve as test oracles by checking whether a program that passes type
checking incurs a memory bug during execution.
However, since type checking is outside their scope, they cannot identify
soundness bugs that do not manifest as undefined behavior.

To empirically validate their effectiveness, we applied them to the code from
the eleven soundness issues that result in memory bugs.
In our experiments, both AddressSanitizer and Miri successfully detected memory
bugs in all cases.
Therefore, using AddressSanitizer or Miri as a test oracle is a promising
direction, although complementary tools are necessary to cover a broader range
of soundness bugs.

\subsubsection{Chalk}

Chalk~\cite{gh-chalk} is an experimental trait solver that serves as a
prototyping ground for designing and improving the trait solver in \rustc{}.
It determines whether certain \C{impl}s are overlapping or orphan and whether a
given type implements a specific trait.
Chalk approaches these tasks in a fundamentally different way from \rustc{};
it models Rust declarations using logical predicates and performs checks by
solving logical formulas.
Thus, it is reasonable to utilize Chalk as a test oracle to discover
trait-related soundness bugs.
However, since Chalk targets only the trait system, other type system features
remain outside its scope.

To assess its effectiveness, we applied the latest version (0.104.0) to the
seven trait-affecting issues (99554, 105787, 106040, 114061, 114389, 126079, and
133361).
%
%
Unfortunately, Chalk currently cannot express Issue 106040 because it does not
support associated types with bounds in \C{impl}s.
For the remaining issues, Chalk correctly identified the orphan \C{impl} in
Issue 99554 and the dyn-compatibility violation in Issue 126079;
it also rejected Issue 105787, but the message incorrectly cited an orphan
\C{impl} rather than overlapping \C{impl}s;
it crashed on Issue 114061 and accepted Issues 114389 and 133361.
Consequently, we conclude that Chalk is yet immature for use as a practical test
oracle.

To facilitate its realistic use as an oracle for trait-related soundness bugs,
we suggest improving its robustness and correctness by testing it with diverse
inputs.
As Chalk targets only the trait system, we do not consider it a general oracle
that can cover all kinds of soundness bugs, and we do not suggest extending it
to cover features irrelevant to traits.

\subsubsection{a-mir-formality}

\begin{figure}[t]
\begin{lstlisting}[caption={Issue 99554 in a-mir-formality}, label={lst:a-mir-formality}]
[ crate a { trait X<ty T, ty U> {} },
  crate b { struct S {}
            trait Y { type A: []; }
            impl<ty T> Y for T { type A = T; }
            impl<ty T> X<S, T> for <T as Y>::A where T: Y {} } ]
\end{lstlisting}
\vspace{-1em}
\end{figure}

The a-mir-formality project~\cite{gh-amirformality} aims to develop an
executable formal model of the Rust type system, which can decide whether a
given program should pass type checking or not.
As its scope is not limited to a specific part of the type system, we believe it
has strong potential to become a comprehensive test oracle for type soundness.

To validate its effectiveness, we applied a-mir-formality at commit \C{ab4f18e}
to the collected soundness issues.
As a-mir-formality defines its own syntax, we manually rewrote the code;
for top-level items, its syntax is close to Rust's;
for function bodies, it follows MiniRust~\cite{gh-minirust}, which resembles the
mid-level intermediate representation (MIR) used in \rustc{}.
For example, \cref{lst:a-mir-formality} shows the code for Issue 99554 rewritten
in a-mir-formality, where a single \emph{program} consists of a list of crates
and a type parameter is introduced with the keyword \C{ty}.
While rewriting the code, we found that it currently supports only a subset of
Rust and MiniRust;
it lacks support for implied bounds, \C{impl} traits, trait objects, function
pointer types, closures, asynchrony, array types, and casts.
Furthermore, the \C{?} operator cannot be used because it exists only at the
surface level and is desugared when code is lowered to MIR.
As a result, we could express code snippets for only five issues (99554, 98117,
106040, 109628, and 114061).
In our experiments, a-mir-formality rejected three (99554, 98117, and 109628)
but accepted the others.
Therefore, we conclude that a-mir-formality is also currently immature for
practical use.

To leverage its strong potential as an oracle, we suggest supporting more
features.
We believe important features to prioritize include implied bounds, \C{impl}
traits, and trait objects.

\subsection{Documentation}

In this section, we describe \emph{holes} in the existing documentation,
focusing primarily on the Rust Reference~\cite{rust-reference} and the
FLS~\cite{fls}, while also referring to Rust RFCs~\cite{rust-rfcs} where
appropriate.
Both the Reference and the FLS describe the syntax in BNF and provide brief
explanations for each syntactic construct.
While the Reference has been maintained as an official Rust project for many
years, the FLS was initially developed by Ferrous Systems and AdaCore for use in
safety-critical industries before being adopted as a Rust project in
2025~\cite{adopting-fls}.
Although their explanations overlap in many areas, the level of detail provided
for specific features often varies between them.
On the other hand, RFCs comprise a collection of proposed changes to the
language.
As discussed below, certain features are explained in detail only within RFCs
even after their stabilization.
While investigating these documents, we reviewed all sections containing
relevant keywords by utilizing the search functionality on their websites,
avoiding incorrectly claiming that certain information is missing.

Note that the purpose of this discussion is to demonstrate that constructing a
mechanized specification for Rust cannot be achieved simply by collecting
existing documents;
rather, it requires significant effort to fill these holes.
Our goal is not to criticize existing documents;
they were not intended to serve as specifications and remain valuable for
providing intuitive understanding.
Since we only aim to demonstrate that the current documentation contains
\emph{some} holes, our discussion does not systematically cover every language
feature.

%

{\it Implied bounds and well-formedness.}
The Reference explains implied bounds by stating that ``lifetime bounds required
for types to be well-formed are sometimes inferred,'' but it does not define
what well-formed types are.
Similarly, the FLS does not provide a precise explanation of implied bounds,
stating only that ``an implied bound is \ldots the byproduct of relations
between lifetime parameters and function parameters \ldots'' without further
clarification on how such relations introduce bounds as a byproduct.
The definition of well-formedness exists only in RFCs, notably RFC 1214
\C{projections-lifetimes-and-wf}~\cite{rfc1214}.
This RFC defines well-formedness using inference rules based on its own syntax
for Rust types.
Unfortunately, this syntax lacks several types present in Rust, most notably
\C{impl} trait types.
Furthermore, some inference rules are unclear;
e.g., one rule includes the premise \C{('x: 'a) in Env}, yet \C{Env} is never
formally defined.

%
%

{\it Coercion.}
In Issue 99158, coercion from a function item to a function pointer breaks type
soundness.
The Reference defines this kind of coercion vaguely, stating only that
``coercion is allowed \ldots [from] function item types to \C{fn} pointers.''
%
%
The explanation in the FLS is more refined, specifying that coercion is
permitted when
``the source's function signature is a subtype of the target's function
signature.''
The FLS further defines the subtyping of function pointer types (presumably
including function item signatures) using the standard notion of variance, where
they are contravariant in their parameter types and covariant in their return
types.
However, this definition does not cover higher-rank cases, where functions have
type or lifetime parameters, possibly accompanied by bounds.
Consequently, it remains unclear whether coercing generic functions to function
pointers requires specific checks for their bounds or permits the substitution
of these parameters with concrete types or lifetimes.

{\it HRTBs and associated types.}
In Issue 141713, the use of the type path \C{F::Output} triggers the problem,
where \C{F} has an HRTB.
We searched the documentation using the keywords ``higher-rank,'' ``associated
type,'' and ``type path'' to determine how an associated type projected from a
type parameter with an HRTB should be interpreted.
Unfortunately, we found explanations only for HRTBs and associated types in
isolation, with no documentation regarding their interaction.
Notably, the description of how a type path is resolved to a type is
insufficient even when HRTBs are not involved.
The Reference defines the syntax of type paths in its ``Paths'' section but does
not explain their resolution process.
The FLS includes an ``Entities and Resolution'' section, which states that \C{<S
as T>::A} ``resolves to the associated type of \C{S}'s trait implementation of
trait \C{T}.''
However, it provides no information regarding the resolution of type paths that
lack such explicit trait disambiguation.




\section{Threats to Validity}
\label{sec:validity}
The primary threat to external validity resides in the selection criteria and
the representativeness of the collected issues.
We are certain that all 30 issues are indeed soundness issues because we studied
each in depth;
readers can also validate this claim by reviewing the code snippets in our
supplementary material~\cite{supp} or the issues themselves on GitHub, as the
dataset size is manageable.
However, additional soundness issues may exist that were missed due to our
conservative selection criteria.
The automatic filtering process relied on GitHub labels;
although these labels are routinely assigned by \rustc{} developers, some issues
might not have been properly labeled at the time of collection.
While we followed the classification used in prior work~\cite{liu2025empiricala}
to identify labels related to type checking, other relevant labels might have
been overlooked.
Moreover, we filtered out issues with labels that appeared irrelevant to
soundness, but it is possible that a soundness issue also revealed other
problems and was thus excluded.
Notably, as mentioned in \cref{sec:method}, 7 issues in the dataset of the
previous work were excluded by our filtering due to their labels, confirming the
risk of missing soundness issues.
Even though we complemented our dataset with these 7 specific issues, other
soundness issues might have been missed for the same reason.
Finally, our manual scan retained issues only if they provided clear evidence
that the expected behavior was rejection, which could exclude less-documented
cases.
Nevertheless, even if the dataset is not exhaustive, we believe our findings
remain valid as they are derived from actual soundness issues.

Regarding this threat, we conducted a lightweight study of issues excluded due
to the I-ICE label, which indicates that the compiler crashes because of an
internal compiler error (ICE).
We found that 181 excluded issues have the I-ICE label, while only one of them
also has the I-unsound label.
As a soundness issue is not necessarily marked with I-unsound, this does not
strictly mean that all except one are not soundness issues.
That said, we believe this result suggests that most I-ICE issues are unlikely
to be soundness issues.
The primary reason is that an ICE prevents code from compiling, while soundness
issues require code that should not type-check to compile successfully.

Another threat to external validity stems from the exclusion of issues related
to feature-gated features.
Inspecting such issues would reveal additional soundness issues, especially
considering that unstable features likely have more compiler bugs than stable
features.
We found that among 223 issues excluded due to an ``F-'' label, 10 issues also
have I-unsound, corroborating this expectation.
However, we believe that focusing on soundness issues within stable features
helps better understand the severity of soundness issues in \rustc{}.

A threat to construct validity concerns the accuracy of our characterization of
features, symptoms, consequences, and bug status.
The identification of these attributes is inherently susceptible to researcher
bias or human error.
To mitigate this, we employed three strategies.
First, two authors independently investigated the issues and conducted extensive
discussions to reach a consensus on the characterization of each issue.
Second, we cross-referenced each issue with multiple external resources,
including related issues, PRs, and discussions on the official Rust Zulip
channels, as well as documents such as the Reference, the FLS, and RFCs.
Third, we designed objective criteria for bug status and checked the status by
type-checking each code using the latest \rustc{} release and inspecting issue
labels and their assigners.

Finally, a threat to conclusion validity lies in the small size of the dataset,
which limits the generalizability of our findings.
Nevertheless, we believe our work provides value to the community.
Our main contribution is reporting that \rustc{} has soundness bugs despite
Rust's emphasis on type soundness, highlighting the urgent need for research
community efforts.
Our in-depth, multidimensional analysis yields actionable insights that
facilitate future research on testing the Rust type checker and formalizing the
type system.

\section{Related Work}
\label{sec:related}
\subsection{Studying Compiler Bugs}

Previous work has investigated bugs in Rust compilers, analyzing a wide range of
issues; in contrast, this work focuses specifically on soundness bugs.
Xia et al.~\cite{xia2023rustcompilers} analyzed historical bugs in two Rust
compilers, \rustc{} and \C{rust-gcc}~\cite{rust-gcc}.
They identified type checking as the most bug-prone component in both compilers,
along with method resolution and IR processing.
Yu and Wang~\cite{yu2024empiricalrustpl} examined bug reports and their
corresponding revisions in the \rustc{} GitHub repository.
They found that traits and ownership are the primary Rust-specific features
involved in compiler bugs.
Their analysis of bug symptoms revealed that internal compiler errors (ICEs) are
the most prevalent, with soundness issues ranking as the second most frequent
occurrence.
Liu et al.~\cite{liu2025empiricala} investigated \rustc{} bugs across various
compiler components.
Their study identified type system errors as the leading cause of bugs, with
errors related to trait bounds and opaque types representing the largest share.
Furthermore, they found that a significant portion of bugs are correctness
issues, including completeness and soundness issues, which constitute the most
frequent symptom after crashes.

Researchers have also studied compiler bugs in domains other than Rust.
Notably, they have covered GCC and LLVM~\cite{sun2016gccllvm},
WebAssembly~\cite{romano2021wasm}, JVM-targeting languages, including Java,
Scala, Kotlin, and Groovy~\cite{chaliasos2021jvm},
Solidity~\cite{li2025solidity}, and deep learning compilers~\cite{du2021dl,
shen2021dlcompilerbugs}.

A notable previous study discovered new compiler bugs instead of analyzing bugs
from programmers.
Amin and Tate~\cite{amin2016java} presented soundness bugs in Java and Scala,
where they utilized null pointers with existential types to enable casting
between arbitrary types without relying on built-in downcasting.
They noted that the identified unsoundness results from the interaction of
different language features, aligning with our findings.

\subsection{Formalizing the Rust Type System}

While this work investigates existing documents, including the Reference, the
FLS, and RFCs, researchers have formalized the type system for various subsets
of Rust, which may aid the construction of a mechanized specification in the
future.
Jung et al.~\cite{jung2017rustbelt} proposed RustBelt, formalizing Rust's
ownership and borrowing type system by defining the core language
$\lambda_{Rust}$ and proved its type soundness.
Wagner et al.~\cite{wagner2025linearity} proposed Borrow Calculus, which
incorporates linearity and borrowing, and proved its soundness.
Additionally, Milewski~\cite{milewski2015traits} formalized the Rust trait
system by defining subsets of the language that include associated types, trait
objects, and coherence rules, and proved type soundness.
However, none of these formalizations can represent any of the soundness issues
studied in this work.
RustBelt and Borrow Calculus provide variants of lambda calculus that express
lifetimes but lack many high-level features, including implied bounds and
traits;
Milewski's languages focus only on traits, omitting lifetimes, and define
coherence rules only for limited language features, without encompassing
associated types or trait objects.
While formalizing all language features would not be feasible, we suggest that
covering implied bounds and treating lifetimes and traits simultaneously should
have the highest priority.

\section{Conclusion}
\label{conclusion}
This work studies 30 soundness issues in \rustc{}, analyzing their affected
features, symptoms, consequences, triggering features, community consensus, and
lifecycles.
We also investigate potential test oracles, such as AddressSanitizer, Miri,
Chalk, and a-mir-formality, and documents describing the language semantics,
such as the Rust Reference, the FLS, and Rust RFCs.
We found that soundness issues are often triggered by associated types and the
interaction between lifetimes and traits, while implied bounds and trait objects
are especially relevant to memory bugs.
Existing tools are insufficient to serve as comprehensive test oracles, and a
huge gap remains between existing documentation and the requirements for a
mechanized specification.

\section*{Acknowledgements}
The Python scripts for the data collection in \cref{sec:method} and the visualization of \cref{fig:timing} were developed with assistance from Gemini 3 Flash.
This material is based upon work supported in part by
the Defense Advanced Research Projects Agency (DARPA) under Agreement No.
HR00112590130,
the National Research Foundation of Korea (NRF) (2021R1A5A1021944,
2022R1A2C2003660, and 2026-25549584), and
Institute of Information \& Communications Technology Planning \& Evaluation
(IITP) grant funded by the Korea government (MSIT) (2024-00337703).
Any opinions, findings, and conclusions or recommendations expressed in this
material are those of the authors and do not necessarily reflect the views of
the funding agencies.

\section*{Data Availability}
The dataset and source code used for this study are available at \url{https://doi.org/10.5281/zenodo.20728406}~\cite{artifact}.

\bibliographystyle{ACM-Reference-Format}
\bibliography{ref}

\end{document}